\documentclass[a4paper, 10pt]{article}
\newlength\aftertitskip     \newlength\beforetitskip
\newlength\interauthorskip  \newlength\aftermaketitskip

\usepackage{rotating}
\makeatletter
\def\maketitle{\par
 \begingroup
   \def\thefootnote{\fnsymbol{footnote}}
   \def\@makefnmark{\hbox to 0pt{$^{\@thefnmark}$\hss}}
   \@maketitle \@thanks
 \endgroup
\setcounter{footnote}{0}
 \let\maketitle\relax \let\@maketitle\relax
 \gdef\@thanks{}\gdef\@author{}\gdef\@title{}\let\thanks\relax}
\makeatother

\makeatletter
\def\@startauthor{\noindent \normalsize\bf}
\def\@endauthor{}
\def\@starteditor{\noindent \small {\bf Editor:~}}
\def\@endeditor{\normalsize}
\def\@maketitle{\vbox{\hsize\textwidth
 \linewidth\hsize \vskip \beforetitskip
 {\begin{center} \Large\bf \@title \par \end{center}} \vskip \aftertitskip
 {\def\and{\unskip\enspace{\rm and}\enspace}%
  \def\addr{\small\it}%
  \def\email{\hfill\small\sc}%
  \def\name{\normalsize\bf}%
  \def\AND{\@endauthor\rm\hss \vskip \interauthorskip \@startauthor}
  \@startauthor \@author \@endauthor}}}
  \makeatother
\usepackage{setspace}

\author{\name Jared Amani Greathouse \email jgreathouse3@student.gsu.edu \\
       \addr Department of Public Management and Policy\\
       Georgia State University\\
       55 Park Pl NE, Atlanta, GA 30303, USA
       \AND
       \name Mani Bayani \email mani.bayani@columbia.edu \\
       \addr Economist, Keystone Strategy\\
       New York, New York, USA
       \AND
       \name Jason Coupet \email jcoupet@gsu.edu \\
       \addr Department of Public Management and Policy\\
       Georgia State University\\
       55 Park Pl NE, Atlanta, GA 30303, USA}

\title{Splash! Robustifying Donor Pools for Policy Studies}

\makeatletter

\@twosidetrue \@mparswitchtrue \def\ds@draft{\overfullrule 5pt}
\makeatother
\usepackage{hyperref}
\usepackage{amsfonts}
\usepackage{optidef}
\usepackage{amsmath,bm,mathtools,upgreek,nccmath,amssymb,amsthm}

\newtheorem{assu}{Assumption}

\newtheoremstyle{named}{}{}{\itshape}{}{\bfseries}{.}{.5em}{\thmnote{#3's }#1}
\theoremstyle{named}

\usepackage{algorithm}
\usepackage{algpseudocode}
\usepackage{listings}
\usepackage{xcolor}

\definecolor{codegreen}{rgb}{0,0.6,0}
\definecolor{codegray}{rgb}{0.5,0.5,0.5}
\definecolor{codepurple}{rgb}{0.58,0,0.82}
\definecolor{backcolour}{rgb}{0.95,0.95,0.92}

\lstdefinestyle{mystyle}{
    backgroundcolor=\color{backcolour},   
    commentstyle=\color{codegreen},
    keywordstyle=\color{magenta},
    numberstyle=\tiny\color{codegray},
    stringstyle=\color{codepurple},
    basicstyle=\ttfamily\footnotesize,
    breakatwhitespace=false,         
    breaklines=true,                 
    captionpos=b,                    
    keepspaces=true,                 
    numbers=left,                    
    numbersep=5pt,                  
    showspaces=false,                
    showstringspaces=false,
    showtabs=false,                  
    tabsize=2
}

\usepackage{graphicx} 
\usepackage[utf8]{inputenc} 
\usepackage[style=apa, backend=biber,natbib]{biblatex}
\DeclareLanguageMapping{english}{english-apa}

\usepackage[
        noabbrev,
        capitalise,
        nameinlink,
    ]
    {cleveref}
\crefname{assu}{Assumption}{assumptions}

 \usepackage{setspace}
\usepackage{booktabs} 
\usepackage{array, doi,hyperref} 
\usepackage{paralist} 
\usepackage{verbatim} 
\usepackage{subfig} 

\usepackage{fancyhdr} 

\date{}

\usepackage[OT1]{fontenc}
\begin{document}
\maketitle

\begin{abstract}
Policy researchers using synthetic control methods typically choose a donor pool in part by using policy domain expertise so the untreated units are most like the treated unit in the pre-intervention period. This potentially leaves estimation open to biases, especially when researchers have many potential donors. We compare how functional principal component analysis synthetic control, forward-selection, and the original synthetic control method select donors. To do this, we use Gaussian Process simulations as well as policy case studies from West German Reunification, a hotel moratorium in Barcelona, and a sugar-sweetened beverage tax in San Francisco. We then summarize the implications for policy research and provide  avenues for future work.
\end{abstract}
\section{Introduction}
Twenty years ago, synthetic control methods (SC/SCM) were developed by \cite{Abadie2003} for causal inference in public policy/economic settings as an extension of the difference-in-differences method (DiD). DiD and other methods commonly make some variant of the parallel trends assumption for imputing the counterfactual. Parallel trends posits the average of the untreated units is a good proxy for the counterfactual of the treated unit. However, this assumption may be hard to defend in may empirical settings \citep{augdd}. Thus, SCM posits a modification to DiD: that a weighted average of \textit{certain} untreated unit (donors) is a better proxy for the counterfactual than an average of all of them. The original synthetic control method (OSC) does this via convex hull restrictions, which prevents extrapolation bias and encourages a sparse set of donors getting weight \citep{Abadie2015,Abadie2021}. This makes SCM to be more transparent in quantifying which units contribute to the counterfactual. For this reason and others, SCM has grown in use as a tool for policy analysts \citep{scotlandsc,Abadie2015, Calderon2015,Cole2020,Marinello2021,Bulle2022,Cohn2022, bullinger2021child,sun2019money}. 
\subsection{Problem Statement}
This study is intended to aid policy analysts who are using SCM in the selection of a donor pool. The validity of SCM is predicated on the donor pool (or the set of untreated units) used. The donor pool should ideally be as similar as possible to the treated unit in the pre-intervention period. These units should also not have undergone similar shocks/interventions. This way, policy analysts can be more confident that any deviation from the observed series is the result of the policy change instead of other confounders. But while the \textit{not treated} condition can be fairly straightforward, composing a donor pool that is meaningfully similar to the treated unit, particularly in a policy context, usually involves some subjective researchers decisions. For example, suppose a school institutes a new mathematics curriculum: should donors be restricted based on socioeconomic status– if so, what should the threshold be (i.e., including only communities with median annual incomes greater than \$70,000 a year)? If so, what in practical terms makes this cutoff so different from \$80,000 a year? Should we only use public schools? The same district? Should we only include schools in the same state, bordering states, or even districts in bordering counties from adjacent states? What if a mix of donor types (i.e., states to regions) made better sense as suggested by \cite{SHI2023512}? Should we impose size restrictions,
and if so, what would it even mean empirically to be similarly sized? In the West Germany case for example, \cite{Abadie2015} note that interpolation can occur ``especially if the donor pool contains units with characteristics that are very different from those of the [treated] unit", and thus opted to restrict their donor pool to 16 similar nations to West Germany. \par In addition to questions about similarity, SCMs might be theoretically invalid when we have \(N>>T\) \citep{Abadie2021a,Amjad2018, SHI2023512,mcclelland2022update} because they may cause interpolation biases. \cite{Abadie2022} note that the risk of interpolation bias is high when \(N>>T\) because the SCM may fit to idiosyncratic values of the donor time series. Interpolation is also possible when we have outlier units or substantial volatility of the time series. Particularly in a modern world where policy analysts increasingly leverage large \(N\) datasets with more granularity than countries or states, researchers more frequently encounter settings where interpolation is a much bigger threat, given the levels of dimensionality \citep{Yang2021,Alonso2020,He2022,ocp,yu2022revitalising,dave2022political,scotlandsc}.\footnote{\cite{Alonso2020} have \(\mathcal{T}_0=9\) to \(J=32\), \cite{Yang2021} have \(\mathcal{T}_0=22\) to \(J=278\), \cite{dave2022political} have \(\mathcal{T}_0=11\) to \(J=150\), \cite{He2022} have \(\mathcal{T}_0=10\) to \(J=327\), \cite{yu2022revitalising} have \(\mathcal{T}_0=15\) to \(100 <J <500\), \cite{ocp} have \(\mathcal{T}_0=18\) to \(J=64\), and \cite{scotlandsc} have \(\mathcal{T}_0=20\) to \(J=345\). We do not suggest that these analyses are invalid, rather, we only highlight recent cases of analysts encountering \(N>T\) settings.  } \par

\begin{figure}[!ht]
\centering
\includegraphics[scale=.6]{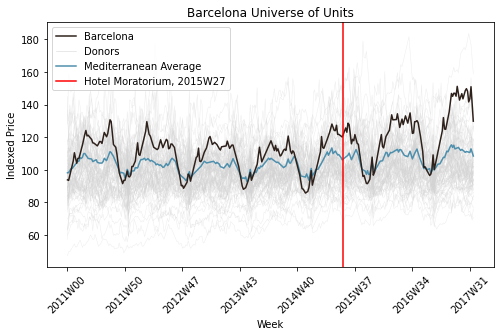}
\caption{Barcelona, 82 City Donors}
 \label{fig:Barcelona}
\end{figure}

For a visual example of a setting where we have noisy outcomes and potentially irrelevant donors, consider \cref{fig:Barcelona} derived from the data of \cite{hotelpaper}, who study the causal impact of Barcelona's hotel moratorium on the daily price of hotel rooms throughout the city. The authors have 82 hotel markets across Europe to compare Barcelona to.
We plot the weekly average of hotel room prices. Even though we have a long pre-intervention time series, it is still clear that many units have prices far higher or lower than Barcelona.  Clearly, we have a complex and noisy time series, with some units exhibiting considerable variance in their weekly prices compared to Mediterranean cities or Barcelona itself. This is a setting where interpolation is possible. By comparison, the original case studies which used SCM-- or the  panel data approach (PDA), SCM's cousin-- had much fewer donors \citep{Abadie2003, Abadie2010, Abadie2015, hsiaopda}.
\par
Whatever the potential source of interpolation bias, the common way of preventing it is to ``judiciously" select a donor pool \citep{Abadie2021}. Thus, a central question for researchers using SCM is ``Which units \textit{belong} in the donor pool in the first place?". Specifically, \cite{Abadie2022} advise analysts to ``trim" their donors to be similar to the values of the treated unit in the pre-intervention period-- and this is where our complications begin. While in practice donor trimming based on outcome/covariate differences is very common \citep{fundment, xinj,olysport,aosla,GRIER2023174}, little formalized guidance for donor trimming exists in the policy context. Thus, trimming is typically based on largely qualitative logic, as different authors will ultimately have different standards on what exactly constitutes `too different' for many given metrics.\footnote{Methods such as trimming via standard deviation distances have been proposed by \cite{mcclelland2022update}, but the author offers no theoretical justification for this metric, and it is unclear how such a procedure would be valid for larger datasets.} \par

However, what if simple trimming is not straightforward: we might ask how to trim down the donor pool from \cref{fig:Barcelona}. What restrictions upon the predictors or other criteria should we use to get a more feasible donor pool-- do restrictions on predictors like climate, unemployment, or geographic factors matter? In \cite{hotelpaper}, the authors have 82 donors, but only a small subset of them are cities located on the Mediterranean Sea. Should we use only Mediterranean donors, or is using cities in France and Spain sufficient? Unfortunately, we cannot even try to answer these questions because we do not know which donors are which-- Booking.com anonymized them to preserve differential privacy. Furthermore, no covariates about the donors were provided. Here, we simply can't make such judgements about what the ``right" set of donors is \textit{a priori}, as there's no predictors to other metrics to start from. Given these constraints (in this example and in many real policy datasets), we argue that leaving the selection of the donor pool to subjective judgment or mere human intuition can compromise the robustness of the results.
\par 
When machine learning experts extended the SCM toolkit, they did so in part to address problems from high-dimensions, missing data, noise, and corrupted data. Some work even addresses interpolation biases \citep{mcclelland2022update, Kellogg2021}, where LASSO or nearest neighbor estimators are used as bias-correction techniques. \cite{Amjad2018} develop a robust SCM, employing singular value decomposition and Bayesian methods to deal with problems such as lack of covariates, missing data, and noisy outcome series, while generating confidence intervals. Later, this approach was extended to a tensor/multiple outcome framework \citep{Amjad2019}. 
\begin{algorithm}[H]
  \caption{Overview of Donor Selection}
  \begin{algorithmic}[1]
    \State Standard Counterfactual Methods in Policy (SCM, DID)
    \begin{itemize}
      \item Find a treatment of interest.
      \item Collect a set of outcomes (and possibly covariates) for each unit.
      \item Drop other treated units/units which underwent shocks to their outcomes.
      \item Trim donors using covariate similarity/domain expertise.
      \item Estimate causal effects via the chosen method.
    \end{itemize}
    
    \State Algorithmic Approaches
    \begin{itemize}
      \item Find a treatment of interest.
      \item Collect a set of outcomes for each unit.
      \item Use \verb|fsPDA| or \verb|fPCA-SYNTH| to select the most relevant donors.
      \item Estimate counterfactuals via constrained OLS or another approach.
    \end{itemize}
  \end{algorithmic}
\end{algorithm}
Even though machine-learning methods have been quintessential in developing new/different objective functions and optimizations for SCM, a curiosity is that not much work exists on formalized methods for donor trimming as a concept.\footnote{Work does exist on covariate selection, for example: One extension by \cite{VivesiBastida2022} employs LASSO to select a sparse set of optimal predictors of the outcome from a large set of irrelevant ones. Statistical learning aside, we also have methods proposed by \cite{cherrypick} to mitigate biases from which covariates to use/pick.} Our paper addresses this gap for public policy analysis. \par
In particular, from the title of our paper, we delve further into how one might further `\textit{robustify}' an SCM estimator. The term comes from \cite{pcr} who emphasize the value of ``first find[ing] a low-dimensional embedding of the data before fitting a prediction model". \cite{pcr} use principal component analysis (PCA) to estimate the low-dimensional embedding of their donors, and then use Ridge regression to impute the counterfactual. We extend the idea of `robustification' when applied to donor trimming: we begin by finding the low-rank structure functional data representation of the pre-intervention outcomes. We then select the most comparable donors to a treated unit. We conclude by fitting a prediction model on the low-rank representation of these most comparable donors. The sketch above highlights how our algorithmic methods to select donors differs from simple use of intuition. \par Specifically, we compare two algorithms: one method we use is functional principal component analysis (\verb|fPCA|) and the other is the forward selection panel data approach (\verb|fsPDA|), both recently presented by \cite{Bayani2021,SHI2023512}.\footnote{In the first draft of this paper, we considered the pairwise deviation minimization method proposed by \cite{yu2022revitalising}. However, this method has no formalized stopping mechanism, which would complicate comparisons. We leave this, and other methods, to future work.} The former is a 5 step method which first estimates the low-rank structure of the pre-intervention time-series and clusters over them to trim down the donor pool, while the latter uses a simple OLS algorithm to pre-process the donors. Both approaches are completely based on pre-intervention outcome trends instead of matching on many predictors, thereby removing one potential source of subjectivity to used to compare the units. Additionally, they are valid in cases where we have many more donors than pre-intervention time periods. Thus, they can alleviate the taxing step of needing to trim down a massive donor pool to a set of relevant units. Our paper is motivated by the following empirical applications:
\begin{itemize}
    \item[--] We revisit and update \cite{Abadie2015}, studying German Reunification. 
    \item[--] We revisit \cite{hotelpaper} which studied the effect of Barcelona's 2015 hotel moratorium on city-wide prices of hotel rooms. How to select the most relevant pool of control units?
    \item[--] Causal impact of sugar sweetened beverage taxes on total employment, revisiting \cite{Marinello2021}.
\end{itemize}
We first compare both trimming methods to OSC in synthetic studies. Then, we implement them in the aforementioned empirical case studies. We presume some familiarity with the SCM and PDA approaches: for a detailed review, see \cite{Hollingsworth2020, scmvpda} or \cite{WAN2018121, Abadie2021}. 
\par
Before we continue, we introduce notation.  Denote scalars, vectors, and matrices as \(x\), \(\mathbf{x}\), and \(\mathbf{X}\). Let \(\mathbb{R}\) denote the real number space. Let \(\|\cdot \|_{1}\) and \(\|\cdot \|_{2}\) denote the vector norms \(L_1\) and \(L_2\) respectively. We are concerned with three sets of units: \(\mathcal{N}\coloneqq \bigcup\{ \widetilde{\mathcal{N}}_{0},\mathcal{N}_{0},\mathcal{N}_{1}\}\). Power set  \(\mathcal{N}\) is the universe of units, indexed by \(j\). Subsets \(\mathcal{N}_{0}\) and \(\mathcal{N}_{1}\) are (un)treated units, defined respectively as \(\mathcal{N}_{0}\coloneqq  \{I+1,\ldots,J\}\) and \(\mathcal{N}_{1}\coloneqq  \{0,\ldots,I\}\). \(\widetilde{\mathcal{N}}_{0}\subset \mathcal{N}_{0}\) is the set of units selected by either \verb|fsPDA| or \verb|fPCA-SYNTH|. We observe a time series for each unit indexed by \(t \in \left(1, T\right) \cap \mathbb{N}\). Partition our time series into sets \(\mathcal{T}\coloneqq \mathcal{T}_{0} \cup \mathcal{T}_{1}\) with their own cardinalities, where \(\mathcal{T}_{0}\coloneqq  \{1\ldots T_0 \}\) is pre-intervention periods and \(\mathcal{T}_{1}\coloneqq \{T_0+1\ldots T \}\) denotes post-intervention periods. We also observe outcomes \(\mathbf{Y}\coloneqq \left (y_{jt}\right)_{j \in \mathcal{N},t \in \mathcal{T}} \in \mathbb{R}^{\left(J+1 \right)\times T }\). Treated unit outcomes are \(\mathbf{y}_0 \coloneqq [y_{0t} \ldots y_{0T}]^{\top}\) and the outcomes for donors are a \(T \times J\) matrix \(\mathbf{Y}_{\mathcal{N}_0} \coloneqq [y_{1t} \ldots y_{JT}]^{\top}\). An indicator \(d \in \{0,1\}\) is the treatment. We assume a linear factor model as a DGP, where \(y^{0}_{jt}\) and \(y^{1}_{jt}\) represent the status of the units being untreated or treated.
\begin{assu}[Potential Outcomes]\label{as:1}
As  described by \cite{porney}
\[
  \mathbf{Y} \coloneqq \left.
  \begin{cases}
    y^{0}_{jt}=\boldsymbol{\mu}_{j}\boldsymbol{\delta}_{t}+\boldsymbol{\theta}_{t}\mathbf{Z}_{j}+\boldsymbol{\epsilon}_{jt} ,\quad d=0 \\
    y^{1}_{jt}=\tau_{jt} + y^{0}_{jt} ,\quad d=1 \\
  \end{cases}
  \right\}
.\]
\end{assu}
Here, \(y^{0}_{jt}\) represents the DGP for all units prior to treatment. If the unit is untreated, we presume the pre-intervention factor model holds. Also, this model prohibits spillover effects of the treatment.  Here \(\boldsymbol{\mu}_{j} = \bigr[\mu_{j1},\ldots \mu_{jr}\bigr]^{\top}\) and \(\boldsymbol{\delta}_{t} = \bigr[\delta_{1t},\ldots \delta_{rt}\bigr]^{\top}\) are \(r \times 1\) factor loadings and common factors, \(\boldsymbol{\theta}_{t}\) are \(1 \times k\) coefficients for \(k \times 1\)  \(\mathbf{Z}_{j}\) predictor effects, and \(\boldsymbol{\epsilon}_{jt}\) represents Gaussian noise \citep[Chapter~8]{Gaillac2023}. The fundamental difficulty of causal inference is that we  observe \(y_{jt} = d_{jt} y_{jt}^1 + (1 - d_{jt}) y_{jt}^0\), where we cannot simultaneously observe a unit being treated and untreated-- thus, our goal is to impute the counterfactual \(\hat{y}_{jt} \). Our treatment effect is simply \(\tau_{jt} \coloneqq  y_{jt}-\hat{y}_{jt}\), where we are typically interested in \(\{\tau_{0T_{0+1}}, \ldots, \tau_{0T_1}\}\). We also make two additional assumptions
\begin{assu}[Existence of Weights]\label{as:2}
\(\exists^{\geq 2} \boldsymbol{\omega}^{\ast} \, \text{s.t.} \, \boldsymbol{\omega}^{\ast}\mathbf{y}_{\mathcal{N}_{0}}^{\prime} \approx \mathbf{y}_{0t \in \mathcal{T}_{0}}\)
\end{assu}

Namely, we presume a \(J \times 1 \) vector of unit weights \(\boldsymbol{\omega}^{\ast} = \left(w_{1} \ldots w_{J} \right) \) will generate a SC, \(\hat{y}_{jt} \coloneqq \boldsymbol{\omega}^{\ast}\mathbf{y}_{\mathcal{N}_{0}}^{\prime}\), such that the counterfactual approximates the observed trajectory of the treated unit in the pre-intervention period. These weights should also match on the common factors which drive the factor loadings from \Cref{as:1}.
\begin{assu}[Factor Conditions]\label{as:3}
There are also factor conditions on variation
\begin{equation}
  \frac{1}{T_0} \sum_{t \in \mathcal{T}_{0}}\boldsymbol{\delta}^{\top}_{t}\boldsymbol{\delta}_{t} \rightarrow \boldsymbol{\Omega} \succ 0  
\end{equation}
and conditions on alpha mixing
\begin{equation}
    \delta(n) = \sup_{A, B} \left| \mathbb{P}(\mathbf{\delta}_t \in A \cap \mathbf{\delta}_{t+n} \in B) - \mathbb{P}(\mathbf{\delta}_t \in A) \mathbb{P}(\mathbf{\delta}_{t+n} \in B) \right|
\end{equation}

\end{assu}

Note that the factor loadings embedded in the donor outcomes are a good proxy to match on the pre-intervention time series of the treated unit [for more on this point, see \citep{Hollingsworth2020}. However, because we do not observe these loadings, we use the time variant outcomes of donors to match on these loadings.  We presume these common factors are alpha mixing, or that there is weak correlation between them such that severe serial correlation won't affect our results, and that there is enough variation in the common factors to match on the unobserved factor loadings. The trimming methods are meant to select some optimal subset of donors, \(\widetilde{\mathcal{N}}_{0} \subset \mathcal{N}_{0}\), such that we maximize the similarities of the donor pool to the treated unit, thereby automating the trimming process. Our causal estimand and diagnostic of interest are:
 \begin{align}
    & \text{ATT} =\frac{1}{\mathcal{T}_{1}}\sum_{j \in \mathcal{N}_{1}} \tau_{jt} & \text{RMSE}=  \sqrt{\frac{1}{\mathcal{T}_{0}}\sum_{t \in \mathcal{T}_{0} } \left(\tau_{jt}\right)^2}.
\end{align}
or, the post-intervention average treatment effect on the treated and the pre-intervention root-mean squared prediction error.
\section{Donor Trimming and Counterfactual Estimation}
First, we review the trimming algorithms. Then, we review the SC and PDA approaches to estimation at a broad level.
\subsection{Forward Selection and Functional PCA}
 Forward selection was recently studied by \citep{SHI2023512, li2022novel} as a viable method of donor trimming for causal inference. Here is a short sketch of it:
\begin{algorithm}[H]
  \caption{Forward Selection}
  \begin{algorithmic}[1]
    \State Initialize $\widetilde{\mathcal{N}}_{0} = \emptyset$.
    \State Initialize $r$ as the iteration index.
    \State Regress \(y_{0t \in \mathcal{T}_{0}}\) upon each \(y_{jt \in \mathcal{T}_{0}}\)
    \State Store \(\widetilde{\mathcal{N}}_{0} \coloneqq j_r \gets \underset{j \in \mathcal{N} \setminus \widetilde{\mathcal{N}}_{0}}{\operatorname{arg\,max}}\, R^{2}(\mathcal{N}_{r-1} \cup \{\hat{j}\}) \).
    \State Update \(\widetilde{\mathcal{N}}_{0}=\mathcal{N}_{r-1}\cup {\hat{j_{r}}}\)
    \State Repeat until \(r>R\).
  \end{algorithmic}
  \label{FS}
\end{algorithm}
where $R$ is the stopping time hyperparameter to be tuned. \citep{ SHI2023512} used a modified BIC \citep{MBICPaper} to estimate this hyperparameter:
 \begin{equation}
\hat{R} = \operatorname*{arg\,min}_{r \in \mathbb{N}} \left\{ \log \left( \hat{\sigma}_{\widetilde{\mathcal{N}}_{0}}^{2} + \log \log N \cdot r \bigl(\textrm{log } T_1 \bigr)/T_1 \right) \right\}.
\end{equation} 

In this method, we basically regress the outcome variable of the treated unit on the outcome variable of all donor members over the pre-intervention time period. We then store the \(R^{2}\) statistics from each estimation, and retain the unit in \(\widetilde{\mathcal{N}}_{0}\) with maximum \(R^{2}\). We then repeat this process again, until there's no significant improvement in the prediction. 
\par

Aside from linear regression or Euclidean distance as advocated by \cite{Abadie2022}, one viable method to extract a donor pool is \(k\)-means clustering \citep{olukanmi2020rethinking}. \(k\)-means clustering finds cluster centers indexed by \(i\) \(\lbrace \mathbf{C} = c_{1},c_{2},\ldots ,c_{k}\rbrace\), defined as

\begin{equation}
    c_{k} = \frac{1}{{\left| {y_{i} } \right|}}\mathop \sum \limits_{{y \in Y_{i} }} y
\end{equation}
such that \({\bigcup }_{i = 1}^{k} y_{j} = \mathbf{Y}_{t \in \mathcal{T}_0}\) and \(y_{i} \cap y_{j} = \emptyset\). Suppose \(\mathbf{G}=[g_{tk}]_{N\times C}\) is an indicator matrix, where \(g_{tk} \in \{0,1\}\) denoting, \(\forall j \in N\) if \(y_{t}\) belongs in a given cluster. The solution to k-means clustering minimizes the within-cluster variance
\begin{equation}
\operatorname*{minimize}_{\mathbf{C}, \mathbf{G}} \Vert \mathbf{Y} - \mathbf{C}\mathbf{G}^{\top} \Vert _{2}^{2}
\end{equation}
Note the maximum number of clusters allowed is a hyperparameter in the K-means algorithm. To avoid manually choosing of the number of clusters, we use Silhouette statistics to ascertain the optimal number of clusters \citep{ROUSSEEUW198753}, computed via
\begin{align} 
s(i) = \begin{cases}
    1 - \frac{b(i)}{w(i)} & \text{if } b(i) < w(i) \\
    0 & \text{if } b(i) = w(i) \\
    \frac{w(i)}{b(i)} - 1 & \text{if } b(i) > w(i)
\end{cases}
\end{align}
After the clustering, we keep the cluster with the treated unit. The new donor pool is \(\widetilde{\mathcal{N}}_{0} = C \setminus \{c_i \mid j = 0\}\), or all of the units in the \(i\)th cluster except for the treated unit. Some papers in recent years have adopted this exact approach to trim donors \citep{LEE2021103617,rudholm2022does}. \par
However, clustering on raw data in policy or economic settings can be very misleading, particularly when dealing with \(N>>T\) or \(T>>N\) settings. The reason for this is that there may be outlier units (as a whole unit or at specific points in time), noise, and many irrelevant units-- what's more, standard clustering over the raw data might detect incidental similarities between units (and cause overfitting). We are interested in the broader temporal correlations between the treated unit and donor units \citep{clushdrev}, and do not wish to fit to noise. So, to address this problem, the first step of \verb|fPCA-SYNTH| is to reduce the data's dimensionality via functional Principal Component Analysis (fPCA) before clustering. \par

One of the main limitations of standard PCA as used in \cite{pcr} is that ``the data points on each curve are assumed to be independent of each other, but in reality it is known that any point on a continuous time-series is correlated with the data points that precede and follow that point'' \citep{fpcasport}. Thus, a main advantage of fPCA over standard PCA is that accounts for temporal correlation within the data points \citep{WARMENHOVEN2021110106} via functional basis splines while reducing the data dimensionality. Hence, the functional PCA scores preserve the temporal correlation among unit observations. Once we have the low-dimensional representation of the outcomes matrix, we may now k-means cluster to extract our donor pool. We would like to highlight some differences between these algorithms: forward selection is a greedy parametric algorithm in that it seeks to maximize the \(R^{2}\) statistic to predict the treated pre-intervention series, and has no method of discarding donors once they've been selected. Thus while forward selection performs regularization, it may suffer from overfitting biases as discussed in \cite{li2022novel}, who solves the overfitting probem by using difference-in-differences. In contrast, \verb|fPCA-SYNTH| is non-parametric in nature. Because the clustering is done over the fPCA scores, there is less of a concern for overfitting.

\subsection{Estimating Synthetic Weights}
Broadly, SCMs and PDAs follow a minimization of the treated unit and the pre-intervention donor matrix
\begin{equation}
   \underbrace{\mathbf{y}_{0}}_{\mathcal{T}_{0} \times 1} \coloneqq \begin{bmatrix}
           y_{01} \\
           y_{02} \\
           \vdots \\
           y_{0T_{0}}
    \end{bmatrix}, \underbrace{\mathbf{Y}_{\mathcal{N}_{0}}}_{\mathcal{T}_{0} \times J} \coloneqq \begin{bmatrix} 
    y_{11} & y_{21} & \dots \\
    \vdots & \ddots & \\
    y_{1T_{0}} &        & y_{JT_{0}} 
    \end{bmatrix} 
\end{equation}
The SC from any selected pool is denoted as \(\hat{y}_{jt} \coloneqq \boldsymbol{\omega}^{\ast}\mathbf{y}_{\widetilde{\mathcal{N}}_{0}}^{\prime}\).  OSC solves
\begin{align*}
    \underset{\omega}{\operatorname{arg\,min}} & \quad ||\mathbf{y}_{0} - \mathbf{Y}_{\mathcal{N}_{0}}\omega_j||_{2}^2 \\
    \text{s.t.} & \quad \Delta^{J-1} = \left\{\boldsymbol{\omega}: \omega_{j} \in \mathbb{I} \, \text{and} \, \| \boldsymbol{\omega} \|_{1} = 1 \right\}
\end{align*}
where the weights lie on the simplex \(\Delta^{J-1}\) and are constrained to be within \(\mathbb{I} \coloneqq [0,1] \subset \mathbb{R}\). Practically, constraints force the weights to not allow predictions outside of the donor pool the user supplies. Oftentimes for OSC, analysts use additional predictors to improve pre-intervention fit, where a matrix of importance weights are computed
\begin{equation}
\hat{\mathbf{V}} = \underset{V \in \mathcal{V}}{\text{arg min}} \left( \mathbf{y}_0 - \mathbf{Y}_{\mathcal{N}_{0}} {\hat{\mathbf{W}}} (\mathbf{V}) \right)^{\prime} \left( \mathbf{y}_0 - \mathbf{Y}_{\mathcal{N}_{0}}{\
\hat{\mathbf{W}}} (\mathbf{V}) \right)
\end{equation}
However, the addition of covariates does not by itself mitigate interpolation biases in the large \(N\) setting \citep{Abadie2021a}, or if there are irrelevant donors/noisy outcomes. \par 

With \verb|fPCA-SYNTH|, the donors are already correlated with the treated unit over time via the clustering. However, we still wish to find a low-rank representation of the donor-pool to estimate the counterfactual with.\footnote{See Appendix A for more detail on low-rank estimation as well as additional technical details.} To do this, we use Robust PCA (RPCA) scores. RPCA scores are obtained via the solution to the following optimization
   \begin{mini}|s|
  {\mathbf{L}, \mathbf{S}}{\Vert \mathbf{L}\Vert_{*} + \lambda\|\mathbf{S}\|_1}{}{}
  \addConstraint{\mathbf{Y} = \mathbf{L} + \mathbf{S}}.
 \end{mini}
 where \(\mathbf{L}\) and \(\mathbf{S}\) are signal and noise matrices of the pre-intervention time-series. \(\mathbf{L}\) and \(\mathbf{S}\) are computed via the Alternating Direction Method of Multipliers with the augmented Lagrangian, where the former matrix emphasizes the pre-intervention temporal patterns and the latter shrinks the influence of the \(\mathbf{S}\) matrix with each iteration. We then use \(\mathbf{L}\) (again in the pre-intervention period) to estimate the unit weights for the SC, as this matrix represents the patterns of the donors in the pre-intervention period. Since the units are already correlated, we no longer need to use the convex hull restriction from OSC. RPCA-SYNTH solves the weights using convex cone restrictions
 \begin{argmini}
  {\omega}{||\mathbf{y}_{0}-\mathbf{L}_{\widetilde{\mathcal{N}}_{0}}\omega_j||^{2}_{2}}{}{}
  \addConstraint{\boldsymbol{\omega}:\omega \in \mathbb{R}_{\geq 0} \quad \forall \, j \in \widetilde{\mathcal{N}}_{0}}.
 \end{argmini}
We expect the estimation of \verb|fPCA-SYNTH| to be less sensitive to outliers and noise compared to OSC, since it does not formally partition the observations using low-rank techniques. Forward Selection PDA estimates the counterfactual as follows using OLS.
 \begin{argmini}
  {\omega}{||\mathbf{y}_{0}-\mathbf{Y}_{\widetilde{\mathcal{N}}_{0}}\omega_j||^{2}_{2}}{}{}
  \addConstraint{\boldsymbol{\omega}:\omega \in \mathbb{R} \quad \forall \, j \in \widetilde{\mathcal{N}}_{0}}.
 \end{argmini}
It uses its optimal donors as obtained from its donor trimming algorithm.

\section{Synthetic Studies}
The goal of the simulations and empirical analyses are to illustrate the utility of algorithmic donor trimming. Thus, while we also study the treatment effect and pre-treatment fit, our simulations have the additional demand of picking the optimal donor pool. We meet this challenge with Gaussian Processes (GP).  GPs define a distribution over functions. More formally, GPs are collections of random variables \(\mathbf{x} \in \mathbb{R}^d\) where a finite set \(p(f(\mathbf{x}_1),\ldots,f(\mathbf{x}_N))\) is derived from a joint Gaussian distribution \(p(\mathbf{f} \lvert \mathbf{x}) = \mathcal{N}(\mathbf{f} \lvert \boldsymbol\mu, \mathbf{K})\). A GP is completely specified by its mean function $m(x)$ (defining its time trend) and covariance function $k(x, x')$ (defining temporal shape/smoothness). $x$ and $x'$ are points in the input space, denoted as \(f(x) \sim \mathcal{GP}(m(x),k(x,x'))\). GPs are versatile approaches to generating complex panel data distributions which might involve non-linear relationships and non-stationary patterns. The final set of processes/donor pools looks like
\begin{align}
\begin{bmatrix} GP(x_1) \\ \vdots \\ GP(x_n)\end{bmatrix}  \sim N\left(\ \begin{bmatrix} m(x_1)\\ \vdots \\ m(x_n)  \end{bmatrix}\ ,\ \begin{bmatrix}k(x_1, x_1)& \cdots & k(x_1, x_m)\\ \vdots & \cdots & \vdots \\ k(x_m, x_1) & \cdots & k(x_m, x_m)   \end{bmatrix}      \ \right).
\end{align}
\begin{figure}[!ht]
\centering
\includegraphics[scale=.8]{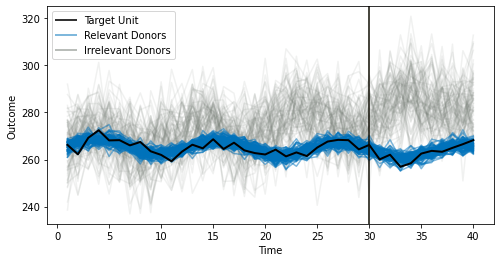}
\caption{Graph of Units}
 \label{fig:SimResults}
\end{figure}
\par Kernel functions are used to define the covariance function in GPs. The first is the Radial Basis Function (RBF) Kernel: \(k(x, x^{\prime})= \text{exp}\left(- ||x - x^{\prime}||^2 \mathbin{/}2l^2 \right)\), to define the mean. The second is the Constant Kernel which scales the overall amplitude of the function \( k(x, x') = \kappa\). We use the Exponential Sine Squared Kernel to model the periodicity in the time series: \(k(x, x^{\prime}) = \text{exp}\left(-2 * \text{sin}^2\left(\left[\pi ||x - x^{\prime}|| \mathbin{/}p \right]\mathbin{/}l^2\right)\right)\). The white kernel models the white noise, denoted as: \( \sigma_n^2 \text{ if } x \equiv x^{\prime}, 0 \mbox{ otherwise.}\) \par
We generate a synthetic high-dimensional dataset of 160 units. We generate two donor pools, each with 80 units per pool. In the treated pool, we select one unit to be treated. \cref{fig:SimResults} shows the treated pool of units (the blue pool) and a set of much noisier, irrelevant donors (the grey pool). We generate these data across 40 time periods, with the hypothetical intervention occurring at \(t=30\). For simplicity, we do not assign a treatment effect, or an ATT of 0 in this case. We report the pre and post-intervention Root-Mean Squared Errors for each algorithm, and discuss the donors selected by each approach.

\begin{figure}[!ht]
\centering
\includegraphics[scale=.55]{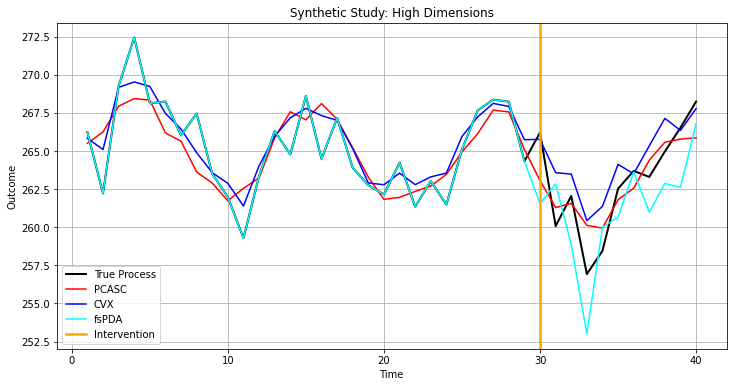}
\caption{Results of Synthetic Studies}
 \label{fig:SCMResults}
\end{figure}
The fPCA method finds 112 units to be the most relevant to the treated unit. The SC is constructed of only 5 units (all of which are from the relevant data generating process). The forward selection method performs much worse, selecting 23 donors, 9 of which are from the unrelated donor pool. OSC
 selects 11 units, 5 of which are irrelevant. fsPDA and the original synthetic control method suggest ATTs of 1.7 and -1.4 respectively. \verb|fPCA-SYNTH| has an ATT of 0.093 (the lowest in absolute value), even though it fits the worst of the methods(1.9 vs 1.4 and 0 for OSC and fsPDA, respectively). \par 
 Following \cite{Amjad2018}, we may use the post-intervention RMSE as a metric for forecasting quality. Precisely, we take the ratio of the pre and post RMSE for each method
\begin{equation}
  Ratio = \frac{
    \sum_{t= T_0+1}^{T} \left( y_{jt} - \hat{y}_{jt} \right)^2
}{
    \sum_{t=1}^{T_0} \left( y_{jt} - \hat{y}_{jt} \right)^2
}  
\end{equation}

 The ratio for \verb|fPCA-SYNTH| is 0.92, for OSC is 3, and for forward selection is 566. Of these, fsPDA has the highest ratio because it clearly overfits the preintervention period, so as a result we have in poor out of sample performance. OSC has the next highest (also having the second best pre-intervention fit) ratio, suggesting that it also overfits to the pre-intervention period. The simulation results highlight the benefits of preprocessing and denoising: in situations where there are linear trends (as we use here), noise, and many donors to choose from, the OSC won't match on the units which are most representative of the target unit's data generating process. Having methods which are robust to outliers, trendy time series, and noise are of use to policymakers because they offer safeguards against overfitting biases which can draw the wrong conclusions.

\section{Empirical Application}
\subsection{West Germany}
First we revisit the West Germany example by \cite{Abadie2015}, who study how West Germany reunifying with its Eastern counterpart affected the GDP per capita of West Germany. Before we continue though, we must lay some essential ground rules for this exercise: this study will \textit{not} be a perfect replication of \cite{Abadie2015} simply because the exact outcome they use (as far as we are aware) is not available.\cite{Abadie2015} use ``GDP per Capita (PPP, 2002 USD)" from OECD National Account statistics. As far as we know, the only data OECD keeps now for GDP per Capita is for 2011 dollars. In absence of the original dataset with all of the excluded donors, we use the newest data available from the Madison Project \citep{bolt2020maddison}. \par The authors begin with 23 OECD nations, discarding Luxembourg and Iceland due to their small and `peculiar' economies. They also discard Turkey due to small GDP per Capita relative to the other OECD nations. Finally, the authors discard Ireland, Finland, Sweden, and Canada as they all experienced structural shocks to their economies in the 1990s, finally limiting their donor pool to 16 OECD nations. \cite{Abadie2015} write, however, their results are robust to the inclusion of Ireland, Finland, Sweden, and Canada. So, we replicate these analyses using the original 16 donors, while also adding in the 4 nations \cite{Abadie2015} excluded (we even add Turkey), giving us 21 donors.\footnote{Note that our results, with the exception of using all 23 donors, are also robust to including either Iceland or Luxembourg (in addition to Turkey).} We do not include predictors in the OSC estimator (the original paper used seven predictors).\par
\begin{figure}[!ht]
\centering
\includegraphics[scale=.55]{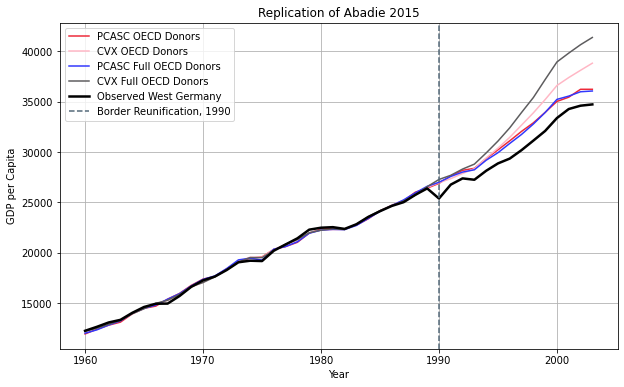}
\caption{Germany Replication}
 \label{fig:Germanresults}
\end{figure}

 
As before, we first estimate the counterfactual using \verb|fPCA-SYNTH| and OSC (we omit the forward selection algorithm for clarity). \cref{fig:Germanresults} plots the observed trajectory of West Germany, and we overlay the four counterfactuals we estimate. In the plot, ``OECD Donors" refers to the original 16 donors used by \cite{Abadie2015}. The ``Full OECD Donors" means Ireland, Finland, Sweden, Canada, and Turkey were included.

\begin{sidewaystable}
\begin{tabular}{l|lrllllrlllll}
\hline
 Pool       & Method   &   NOR & FR   & AUSTR   & UK   & JP   &   NL & AUS   & PORT   & TURK   & US   & GRC   \\
\hline
 OECD Donors      & PCASC    &    0.277 &          &           &                  &         &         0.557 & 0.122       &            &          &                 &          \\
 OECD Donors      & CVX      &    0.089 & 0.12     & 0.282     & 0.051            & 0.167   &         0.115 & 0.039       & 0.078      &          &                 & 0.058    \\
 Full OECD Donors & PCASC    &    0.177 &          & 0.142     & 0.029            &         &         0.366 &             &            &          & 0.195           &          \\
 Full OECD Donors & CVX      &    0.091 & 0.002    & 0.294     &                  & 0.2     &         0.091 &             &            & 0.17     &                 & 0.152    \\
\hline
\end{tabular}
 \caption{\label{tablefirst}Germany Weights}
\end{sidewaystable}
We can see that the weights selected by \verb|fPCA-SYNTH| are Australia, Netherlands and Norway [all of which were also selected into the donor pool in \cite{Bayani2021}]. When we include all of the donors, it weighs Austria, Netherlands, Norway, United Kingdom, and United States. Notice that all of these with the exception of the United States were also included as donors in \cite{Bayani2021}. The imputed counterfactual is also robust to the exclusion of the United States. OSC paints a different picture: substantively, the results also suggest that the GDP per capita would otherwise be higher, however, the effect sizes are double those of \verb|fPCA-SYNTH|. Furthermore, the OSC results are sensitive to the inclusion of the other irrelevant donors, even matching on Turkey (which \cite{Abadie2015} note has a particularly low GDP per Capita). While it's true that \verb|fPCA-SYNTH| includes the United States, RPCA algorithm used by \verb|fPCA-SYNTH| guards against the drastic changes in effect sizes.
The results, even in medium dimensions, illustrate the utility of a donor trimming algorithm. OSC is sensitive to changes in the donor pool, and \verb|fPCA-SYNTH| is less so, even getting the same effect size \textit{while} including the United States, which was not included in the cluster in \cite{Bayani2021}.
\subsection{Barcelona}
In July 2015, the city of Barcelona, Spain implemented a moratorium for the building of new hotels, ostensibly because over-tourism had deleterious social and economic impacts to the surrounding environment. \cite{hotelpaper} study the counterfactual of Barcelona's average prices for hotel rooms, using daily pre-intervention data from January 2011 to June 30th 2015 to construct the counterfactual. Here, we estimate the causal impacts of Barcelona's hotel moratorium. It is tempting to follow \cite{Abadie2015} and restrict our donors to Mediterranean cities. However, Barcelona's average price is almost always higher than the average of Mediterranean donors, so it is possible (if not likely) that Barcelona could be \textit{better} approximated by some non-Mediterranean donors. In other words, it is an arbitrary assumption we make that only Mediterranean cities would be the ideal donor set- given how cities like London and others are in the donor pool, we elect to use the universe of units instead of simply limiting our units to the Mediterranean.
\par Our outcome is the mean price of hotel rooms in each city, normalized to 100. Here, \(j=0\) is Barcelona and \(\mathcal{N}_{0}=82 \) represents the universe of donors. While data from \cite{hotelpaper} are in daily format, we use the average weekly price as our outcome. The observation window is from week 1 of January 2011 to week 31 (August 7th) 2017. 

\begin{figure}[!ht]
\centering
\includegraphics[scale=.5]{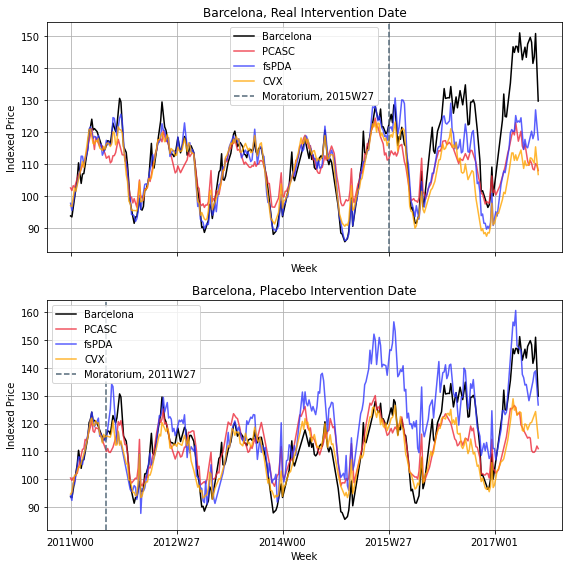}
\caption{Barcelona, Real and Placebo}
 \label{fig:Barcelonaresults}
\end{figure}
After applying fPCA to all pre-intervention weeks, the cluster for Barcelona retained 14 donors, or \(\widetilde{\mathcal{N}}_{0}=13\). Of this subset, 9 donors, or 64\%, were Mediterranean cities. For the weights, 4 donor units were given weight. Of these, 3 were from the Mediterranean Sea donor pool. For \verb|fs|, 11 donors were chosen. Of those 11, 8 were Mediterranean. For OSC, 3 of the 7 donors were Mediterranean. Notice how these methods, particularly the PCA and fs method, appeal  very nicely to intuition. Since \cite{hotelpaper} already hypothesized Mediterranean Donors were more likely to be similar to Barcelona, the fact fPCA and forward-selection both choose donor pools with majority Mediterranean donors reassures us the methods are picking up on donors most related to Barcelona. Equally, the fact we get this result without explicitly adjusting for other predictors of room prices is doubly reassuring for the validity of both methods. In contrast, while OSC has good fit, most of its donors are not Mediterranean. Because SCM is justified on a latent variable model, the counterfactual must also match on unobserved confounding, instead of simply fitting to the pre-intervention period or covariate balance. Since we would imagine that at least many of the weighed units would be Mediterranean, this suggests the algorithmic approaches are more theoretically justified than OSC in setting where there are many irrelevant units. Note, that all methods used here agree in terms of the direction of the average treatment effects on the treated: the moratorium seems to have raised the prices of hotel rooms, on average.
\begin{table}[htbp]
    \centering
\begin{tabular}{llrrrrr}
\hline
 Method   & Intervention         &   ATT\_per &    ATT &   RMSE &   PostRMSE &   Ratio \\
\hline
 CVX      & Main Intervention    &    13.045 & 15.906 &  3.355 &     24.583 &            7.327 \\
 CVX      & Placebo Intervention &     3.131 &  3.549 &  1.973 &     17.291 &            8.764 \\
 PCASC    & Main Intervention    &    11.372 & 12.45  &  5.295 &     16.84  &            3.18  \\
 PCASC    & Placebo Intervention &     1.225 &  1.371 &  3.346 &     10.184 &            3.044 \\
 fsPDA    & Main Intervention    &     9.137 & 10.047 &  2.614 &     14.081 &            5.387 \\
 fsPDA    & Placebo Intervention &    -5.49  & -6.546 &  1.441 &     11.789 &            8.181 \\
\hline
\end{tabular}
    \caption{\label{table1}Mediterranean Study}
\end{table}
\par
Another important point when evaluating the estimators is robustness to changes in the intervention date, to see how well the SC tracks the observed series. In this case, we change the intervention date to week 27 of 2011, a full 4 years before the intervention. Here, \verb|fPCA-SYNTH| assigns two weights of the 19 units it selects as optimal, one of which is Mediterranean (OSC gets the same unit). OSC's weights in this case are actually 57\% Mediterranean. Noticeably, the forward selection method has the most drastic change in the ATT, with the sign flipping completely, and departing far from the observed time series, meeting it only at the end of the time series. By comparison, \verb|fPCA-SYNTH|'s Post to Pre RMSE ratio changes the least compared to the other two methods. Its synthetic Barcelona tracks the observed Barcelona until the intervention date, even though the effect size differs from the first intervention date. The fact that its predictive capacity in terms of Post/Pre Ratio changes so little, being so close to the start of the time series, while retaining a similar proportion of theoretically relevant donors suggests that it is more robust to changes in the model than the OSC method.
\subsection{Sugar Sweetened Beverage Tax}
Globally, sugar sweetened beverage (SSB) taxes, like other forms of excise taxes, have become popular economic policies which are intended to increase public health benefits without simply banning consumption of certain goods. Industry argues SSB taxes will have negative effects on employment, since now fewere people will be able to work in their warehouses/industries that are predicated on SSB revenue. So, the causal evaluation of SSB taxes on consumption of SSBs and secondary economic impacts has become popular among economists and policy analysts [e.g., \citep{BSCSSB}]. \cite{Marinello2021} study the causal impact of San Francisco's SSB tax on various forms of employment, finding a very small impact on employment levels. We replicate their analyses using the estimators presented so far. \par  The authors limit their donor pool to 62 urban counties around the United States, using SCM to evaluate the causal impact. They also have 62 months of pre-intervention data, with the intervention beginning on January 1, 2018. Precisely, our outcome is the total number of people employed in each city/county equivalent. The full dataset in \cite{Marinello2021} consists of 93 donors. So, we keep all 93 donors. Another key deviation from our analysis is that \textit{unlike} \cite{Marinello2021} who use many covariate predictors, we use none. 
\begin{figure}[!ht]
\centering
\includegraphics[scale=.5]{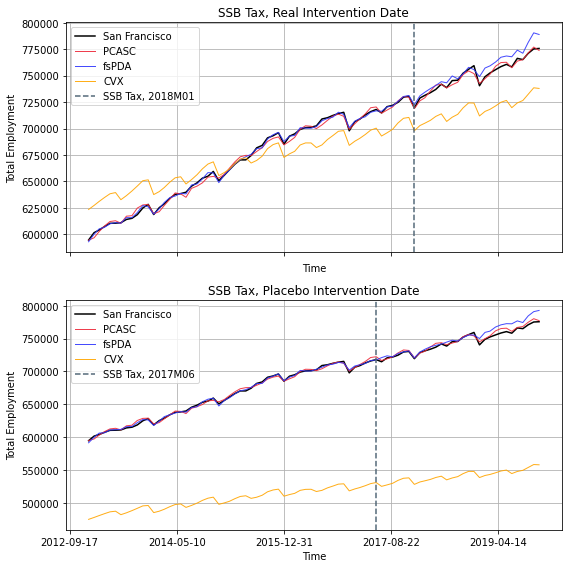}
\caption{Results of SSB Study}
 \label{fig:SSBResults}
\end{figure}

\begin{table}[htbp]
    \centering
\begin{tabular}{llrrrrr}
\hline
 Method   & Intervention         &   ATT\_per &     ATT &    RMSE &   PostRMSE &   Ratio \\
\hline
 CVX      & Main Intervention    &     4.345 &  32.625 &  17.867 &     37.213 &            2.083 \\
 CVX      & Placebo Intervention &    27.376 & 203.821 & 159.161 &    204.755 &            1.286 \\
 PCASC    & Main Intervention    &     0.065 &   0.485 &   2.67  &      2.604 &            0.975 \\
 PCASC    & Placebo Intervention &    -0.27  &  -2.016 &   2.693 &      3.295 &            1.224 \\
 fsPDA    & Main Intervention    &    -0.813 &  -5.647 &   1.39  &      7.061 &            5.08  \\
 fsPDA    & Placebo Intervention &    -0.805 &  -5.591 &   1.651 &      8.043 &            4.872 \\
\hline
\end{tabular}
\caption{\label{table2}SSB Study}
\end{table}
The weights assigned by \verb|fPCA-SYNTH| are: 'Riverside County': 0.096, 'San Benito County': 4.264, 'Denver County': 0.049, 'Hillsborough County': 0.138, 'St. Tammany Parish': 1.924, 'Kings County': 0.171, 'Wake County': 0.122, 'Hays County': 1.005, 'Kendall County': 2.187. For context, Riverside and San Benito counties are located in Combined Statistical Areas in California. Wake, Denver, Hillsborough, and Kings counties are home to Raleigh/Durham, Denver, Tampa, and Seattle, all major major American cities. For forward selection, the OLS coefficients are 'Travis County': 8690.0, 'St. Tammany Parish': 0.788, 'Hartford County': 2.57, 'Baltimore city': -0.715, 'Charles County': 0.71, 'Placer County': 3.42, 'District of Columbia': 0.829, 'Essex County': -0.231. \par Regarding the OSC, we'd like to highlight a mistake we made during the writing of this paper: at first, we'd mistakenly coded the intervention as being in February 2018 instead of January. The pre-intervention RMSE for OSC in this case was 163.725 and the post-intervention error was 208.755. In other words, the line did not fit the pre-intervention period at all. When we corrected this, the pre-RMSE is now 17.867 and post-error is 37.213. While a marked improvement from the specification with the miscoded intervention, we highlight, as others have, how the OSC can be quite sensitive to small tweaks in the model, suggesting its estimates are less robust. Most notably, this also suggests the covariates the authors included originally (7 predictors, 5 county specific and 2 industry specific) can be quintessential to estimating counterfactuals with OSC. For a more realistic in-time placebo test, we report a 6 month placebo test, where we move the intervention to June of 2017 instead of 2018. \par

As we can see, all of the other methods (especially \verb|fPCA-SYNTH|) are quite robust to a longer backdate in the treatment, whereas OSC is not. Note that this does \textit{not} mean the original analyses are wrong or incorrect. Our methods still obtain the main result, that the effect of the SSB tax on employment was quite small. We only echo the comments of others \citep{Bayani2021, Amjad2018} who remark that that the need for covariates to obtain convergence and good fit is a limitation of the original method. Particularly for empirical applications, this need for auxillary predictors can sometimes not be at all trivial. Thus, public policy analysts might prefer methods which both automate the selection of a donor pool while obtaining similar effects. \par 
Now we turn to the estimates with the actual intervention date. \verb|fsPDA| detects a -0.813\% effect, and \verb|fPCA-SYNTH| finding a 0.065\% average impact on employment. The fact that these methods (without predictors) reach qualitatively identical conclusions of the original paper, while at the same time selecting mostly large metropolitan areas as donors suggests that in many cases (such as Washington DC or San Benito County), donor trimming methods or additional advances to the standard SCM may be of use to policy analysts for their robust properties, particularly when there are multiple donors and little covariate information to use for the standard SCM.

\section{Discussion and Conclusion} 
In this study, we delve into the significance of algorithmic methods for selecting the optimal donor pool for SCM. The challenge of accurately identifying an appropriate control group stands at the forefront of causal inference, given that the control group's validity is the basis for the causal studies. The importance of donor pool selection is amplified within the SCM context due to the classic model's vulnerability to interpolation bias, particularly when we have many donors and/or trendy/noisy outcome time-series. Policy analysts oftentimes employ domain expertise or Euclidean distance measures to select donors: our research underscores the necessity of adopting modern machine learning approaches for this task. Put simply, theory is not a substitute for formal algorithmic analysis: if we think that most of the relevant donors should come from a given pool of units, then evidence from unsupervised algorithms should suggest this. 

Particularly as the use of big data to solve policy problems that require causal inference grows \citep{grimmer2015we}, so too should the causal methodologies used to do analysis. Particularly when analysts are faced with many potential donors, algorithmic selection of a donor pool is useful to researchers who use SCMs. We think our results should leave policy analysts and empirical researchers with a few major takeaways. First, researchers can use these methods (and others such as dynamic time-warping) when the research context provides lots of degrees of freedom with regard to donor pool selection, \textit{or} if a particular set of SCM results is sensitive to the composition of the donor pool. So, if leave-one-out analyses are not robust, one viable method would be to use clustering or another method to extract the optimal donors and re-run original SCM analyses. Algorithmic donor trimming reduces the role of arbitrary research decisions. \par   

Our work has limitations, too. Because policy research increasingly substitutes synthetic control for DiD designs \citep{cunninghamsub2023}, having a better understanding of the exact conditions under which an algorithm selects a valid set of donors is useful. In this paper, we use Gaussian Processes and three empirical cases-- however, formal finite sample analyses of the asymptotic properties of k-means clustering or similar algorithms may expand this untapped area of the literature. Given recent finite sample studies of the \(k\)-means algorithms and SCMs \citep{zhang2022asymptotics,ASYMSCM,zeitler2023nonparametric}, more formalized asymptotic study of algorithmic donor trimming methods may be warranted. \par

\printbibliography
\end{document}


\maketitle
\begin{appendix}
\section{Low Rank Methods in SCM}
A low rank matrix is frequently perceived as a matrix that can be closely approximated by a less complex, or compact, matrix. In other words, ``low rankness" might suggest that the data at hand could be effectively represented or approximated by a reduced set of variables or factors, thereby simplifying the underlying structure of the data. In the case of Barcelona for example, while it's true that in a global economy an innumerable amount of factors affect hotel room prices, only a few of them are the most important factors in setting the price (e.g., weather, demand, geography, etc). When we posit that the observed outcomes are low rank, we posit that having access to this general pattern is a better approximation to the real data generating process compared to us simply using the full dataset. \par 

Low rank matrices play a pivotal role across a variety of mathematical and computer science disciplines. Particularly in machine learning, strategies like Principal Component Analysis (PCA) and Singular Value Decomposition (SVD) are regularly employed to decrease the dimensionality of data. However, such methods have not received as much attention by comparison in the public policy setting, at least within the context of the SCM. Dimensionality reduction by PCA or other methods retains the bulk of the crucial characteristics of the data, while simultaneously minimizing the complexity of the dataset. Particularly in settings where complex time series and outliers exist, low-rank estimation has received more attention in economics to extend existing econometric methods. \par

The initial applications of low-rank estimation within SCM were pioneered independently by \cite{MCAthey} and \cite{Amjad2018}. Both models strategically incorporated the low-rank structure into counterfactual estimation, albeit in divergent ways. \cite{MCAthey} deployed low-rank estimation to provide counterfactual assessments for the post-intervention time period, thereby enabling the recovery of the potential outcome's missing values for the treated unit. This paper used the nuclear norm to construct the counterfactual; however, as mentioned by \cite{Bayani2021}, the imputed counterfactual is less interpretable because the method generates no weights to explicitly see which units contribute the most to the counterfactual,  Conversely, \cite{Amjad2018} use the \(L_2\) norm along with SVD to distill the low-rank structure from the pre-intervention donor matrix. While this method does provide explicit weights, they weights are not sparse by design due to the Ridge regression's geometric properties. Either way, this low-rank estimation was then harnessed to find a linear relationship between the donor units and the between the treated unit.Each of these seminal studies has demonstrated distinct yet effective applications of low-rank estimation in SCM studies. \par

The low-rank approximation of a matrix $Y$ can be found using SVD. If $\mathbf{Y}$ is an $M \times N$ matrix, it can be factored using SVD as follows:
\begin{align}
    \mathbf{Y} = \mathbf{U} \boldsymbol{\Sigma} \mathbf{V}^\top
\end{align}
Where: $\mathbf{U}$ is an $M \times M$ orthogonal matrix. $\boldsymbol{\Sigma}$ is an $M \times N$ diagonal matrix and $\mathbf{V}^\top$ is the transpose of an $N \times N$ orthogonal matrix, $\mathbf{V}$. The diagonal elements of $\boldsymbol{\Sigma}$ are singular values of $\mathbf{Y}$. To get a low-rank approximation of $\mathbf{Y}$, we only keep the largest $k$ singular values and correspondingly, $k$ columns of $\mathbf{U}$ and $\mathbf{V}$, thus forming matrices $\mathbf{U}_k$, $\boldsymbol{\Sigma}_k$ and $\mathbf{V}_k$, where $k$ is a scalar denoting the rank of the approximated matrix. Then, the approximation of $\mathbf{Y}$ is:

\begin{align}
  \mathbf{Y}_k = \mathbf{U}_k \boldsymbol{\Sigma}_k \mathbf{V}_k^\top
\end{align}
Here, $\mathbf{Y}_k$ is a low-rank approximation of the matrix $\mathbf{Y}$. 

\par The low rank estimation we employ extends this idea into the realm of donor selection. We seek to obtain a low-rank representation of the time-series in the pre-intervnetion period via use of the fPCA based clustering, and a low-rank representation of the donor matrix using the RPCA method. In the main text, we mention the idea of the linear factor model \(y^{0}_{jt}=\boldsymbol{\mu}_{j}\boldsymbol{\delta}_{t}+\boldsymbol{\theta}_{t}\mathbf{Z}_{j}+\boldsymbol{\epsilon}_{jt}\). As previously stated, this model is fundamentally grounded in the estimation of a low-rank structure. However, it's important to note that this model exhibits a considerable vulnerability to outliers, meaning anomalous or extreme data points may disproportionately influence the overall results. To overcome the issue of outliers, we define a latent variable model as discussed in \cite{Amjad2018}, where our outcomes are generated via
\begin{equation}
    \mathbf{Y}_{jt}= \mathbf{\rho}+ \mathbf{\epsilon}
\end{equation}
where our observed outcomes are a function of \(\mathbf{\rho}\) which is some unknown mean process and error. The mean function may be further decomposed into two components, \(\left(\delta, \theta \right)\)
which follow an unknown structure and may be non-linear. Put differently, a low rank hypothesis posits that for many applications, we may decompose our outcomes into a low-rank component and a noise component. The natural choice to do this is by using RPCA, because it attempts to solve the problem:
  \begin{mini}|s|
  {\mathbf{L},\mathbf{S}}{\text{rank}(\mathbf{L}) + \lambda \Vert \mathbf{S}\Vert_0}{}{}
  \addConstraint{\mathbf{Y} = \mathbf{L} + \mathbf{S}}.
 \end{mini}
 where \(\mathbf{L}\) is the low-rank matrix (which again is the general temporal patterns of the donor units) and \(\mathbf{S}\) refers to the sparse matrix (the noisy component of the observed donor matrix). In other words, we directly extract the latent characteristics of the donor matrix when estimating the weights. However, the solution to this problem is NP-hard. So, we use the surrogate norms to extract these matrices from the RPCA scores
   \begin{mini}|s|
  {\mathbf{L}, \mathbf{S}}{\Vert \mathbf{L}\Vert_{*} + \lambda\|\mathbf{S}\|_1}{}{}
  \addConstraint{\mathbf{Y} = \mathbf{L} + \mathbf{S}}.
 \end{mini}
The RPCA scores themselves are more robust to noise, missing data and severe corruption of data precisely because of the problem breaking the RPCA scores into signal and noise components, which explains the robustness of the approach in the main text.
\printbibliography
\end{appendix}